# Approaching Digital Humanities at university: a cultural challenge

Silvio Peroni, Francesca Tomasi

**Introduction**

In the past twenty years, the University of Bologna has positioned itself as one of the most attractive European centres for research in Digital Humanities (DH). Reaching that status was possible thanks to the effort of a few researchers in the area, often working in different departments. These researchers, dealing with either humanistic studies or technological applications, have dedicated their main research activity to topics related to a domain, i.e. DH, a field of study which still promotes much debate aiming at identifying its disciplinary borders.[1]

At the international level, the definition of DH as a 'big tent',[2] required choosing and defining an exact and well-focused idea of the domain: the reflection on computational models as a way to enhance our cultural context with a consciousness of both the computer science theories, methodologies and technical practices, and the issues, or questions, in the humanities study and research. In detail, this goal was not conceived of as being achieved as the summation of different disciplines. Instead, it had to result in constant collaboration, mutual interchange, cooperation between fields and a final hybridisation of disciplines as a new way of doing research and teaching. In a sentence: it is a new domain of knowledge. Working in the DH means aggregating heterogeneous knowledge from numerous traditional disciplines to propose strategies and ideas and use tools and techniques for research within a new vision of the humanities. Teaching DH means providing the expertise and skills required to convert this knowledge into practice, aiming at approaching humanities through new methodologies and producing innovative outputs.

At a national level, the University of Bologna has been deeply involved in the foundation of crucial DH-related activities and entities since the beginning. Members of the University have been among the founders of the *Associazione per*

*This work has been partially funded by the European Union's Horizon 2020 research and innovation program under grant agreements No 101017452 (OpenAIRE-Nexus) and No 870811 (SPICE).*

[1] The number of publications related to the debate on DH is huge. See for example Alex H. Poole, 'The conceptual ecology of digital humanities', *Journal of Documentation*, 73: 1, January 2007, 91–122. https://doi.org/10.1108/JD-05-2016-0065

[2] 'Big Tent Digital Humanities' is the theme of the Digital Humanities 2011 conference at Stanford University, https://dh2011.stanford.edu/?page_id=97

*l'Informatica Umanistica e la Cultura Digitale* (AIUCD, http://www.aiucd.it),[3] and covered directorship roles during the first decade of the association. As the Italian situation is concerned, AIUCD has been an essential step in recognising the role of people dealing with computing and the Humanities, together with the idea of the existence of a digital culture to promote. With the introduction of a programmatic and organic plan for DH, supported by the political will of the University of Bologna and its Department of Classical Philology and Italian Studies, activities around DH at the University started to flourish. Indeed, in the past five years, scholars created (in chronological order): the International second cycle master's degree in *Digital Humanities and Digital Knowledge* (DHDK, https://corsi.unibo.it/2cycle/DigitalHumanitiesKnowledge),[4] *Digital Humanities Advanced Research Centre* (/DH.arc, http://dharc.unibo.it), and the PhD programme on *Cultural Heritage in the Digital Ecosystem* (CHeDE, https://phd.unibo.it/chede/en).

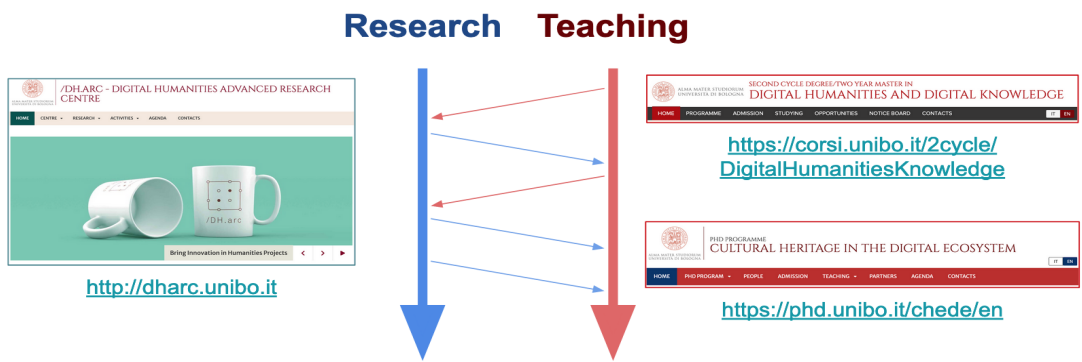

Figure 1. The *research* and *teaching* trajectories implemented within the *plan for DH* at the University of Bologna.

This plan for DH has followed two parallel and complementary trajectories. One concerns teaching DH in proper degrees (master's and PhD) at the University. The aim, in this case, is to shape new professionals acting as a proxy between cultural endeavours and technological expertise, who are highly sought after by national cultural heritage (CH) markets (by CH institutions, libraries, galleries, museums, companies, etc.). The other trajectory is *research*, implemented through the /DH.arc, where scholars have found a stable and fertile common grounding for activating new collaborations and starting novel projects around DH studies.

However, the real added value of this ecosystem lies in the middle. On the one hand, the frequent interactions between the teaching and research trajectories that have been set up over the years have enabled students to increase their skills by measuring their knowledge on real research projects. On the other hand, scholars have taught their courses using groundings and examples from actual and applicative research contexts. The fact that members of the /DH.arc are also the professors teaching in DHDK and members of the PhD committee in CHeDE has simplified the dynamics (and increased the effectiveness) of these interactions.

---

[3] The AIUCD is the Italian Association for Digital Humanities, founded on the 5th of April 2011.
[4] Francesca Tomasi, 'Digital Humanities and Digital Knowledge (DHDK) International second cycle/Master degree', *Umanistica Digitale*, 2, May 2018. https://doi.org/10.6092/issn.2532-8816/7862

This mutual and constant dialogue between research and teaching is also attested by the numerous research projects that students are required to prepare as the assessment method; teachers use this activity to verify the reaching of learning objectives and practical competencies. The projects realised by students, especially websites and code repositories, are collected in an open environment (created, in turn, by a student who won a grant for research projects offered by DHDK). In this environment, each student can create their portfolio (DHDKey, https://projects.dharc.unibo.it/dhdkey/) and has a place for the long-term preservation of research data and applications. These projects represent the link between doing research on a topic using DH methods and techniques, since students are required to elaborate research questions and try to adopt computational models to deal with them, and teaching since projects are tangible outputs of the results of educational practices.

**Informatics from a cultural perspective**

This reform toward implementing a DH ecosystem within the University, accompanied by the recognition that the field has gained at the national level during the past five years, has resulted in the active involvement of several DH students and scholars in national and international projects and activities, which currently are shaping a new cultural heritage system founded on *digital* technologies. However, instead of introducing computational tools as a mere means for enhancing CH studies, the path followed in the University DH courses was to bring back the basics of Informatics (computational thinking,[5] abstraction,[6] algorithms,[7] recursion,[8] data structures,[9] etc.) at the foundation of DH skills.

---

[5] Jeannette M. Wing, 'Computational thinking and thinking about computing', *Philosophical Transactions of the Royal Society A: Mathematical, Physical and Engineering Sciences*, 366: 1881, July 2008, 3717–3725. https://doi.org/10.1098/rsta.2008.0118
[6] Jeff Kramer, 'Is abstraction the key to computing?', *Communications of the ACM*, 50: 4, April 2007, 36–42. https://doi.org/10.1145/1232743.1232745
[7] Robin K. Hill, 'What an Algorithm Is', *Philosophy & Technology*, 29: 1, January 2015, 35–59. https://doi.org/10.1007/s13347-014-0184-5
[8] Jeffrey Watumull, Marc D. Hauser, Ian G. Roberts, & Norbert Hornstein, 'On recursion', *Frontiers in Psychology*, 4, January 2014. https://doi.org/10.3389/fpsyg.2013.01017
[9] Donald E. Knuth, *The Art of Computer Programming, Vol. 1: Fundamental Algorithms, 3rd Edition*, Boston, MA, United States of America: Addison-Wesley, 1997.

It is worth mentioning that the choice of teaching Informatics from this perspective concerns the *cultural* aspect of computing: it does not deal with 'system and tools, but with principles and methods […] focusing on the core scientific concepts of computing, on its conceptual kernel'.[10] Metaphorically, the idea is not to teach how to ride a bike (the tool) but rather to introduce and explain the mechanics (i.e. the theoretical foundations) hidden behind the final product (the bike) to deeply understand how it works, how to fix it in case problems arise or adaptations are needed, and to highlight that several computational tools available are based on identical or similar groundings.

This approach to computing has been implemented by establishing Computer Science courses as one of the three pillars of DH teaching at the University, together with literary, linguistic, historical and cultural courses and other complementary teachings in economy, law and communication.[11]

**Teaching Digital Humanities**

All teaching – independently from the Humanities, Informatics, or complementary settings it may have – is focused on introducing its core material by using examples from real-life experiences. For instance, for the very first course of the DHDK degree, i.e. *Computational Thinking and Programming*,[12] several examples are introduced to students using objects they deal with daily in their experience to recall basic computational concepts: the way the text of a book is organised follows a precise hierarchical structure (a book is made of chapters, each chapter is made of paragraphs, each paragraph is made of text) which is abstracted by the *tree data structure*; the way people are served at the post office is compliant with precise rules of engagement recalling the *first in, first out* (FIFO) strategy which is typical of the *queue data structure*; the infinity mirror configuration recalls the concept of *recursion*,[13] just to name a few examples.

---

[10] Enrico Nardelli, 'Do we really need computational thinking?', *Communications of the ACM*, 62: 2, February 2019, 32–35. https://doi.org/10.1145/3231587

[11] These courses address programming principles and languages, data modelling and multimedia database design, design and creation of web application and interface management, extraction and representation of data, production of multimedia applications in a cultural heritage context, open science.

[12] Marilena Daquino, Silvio Peroni, & Francesca Tomasi, 'Can we do better than computers? Teaching Computational Thinking to Digital Humanists' in Stefano Allegrezza, ed, *AIUCD 2019—Book of Abstracts*, Venice, Italy: Associazione per l'Informatica Umanistica e la Cultura Digitale 2019 (43–48). http://aiucd2019.uniud.it/wp-content/uploads/2020/03/AIUCD2019-BoA_DEF.pdf

[13] Wikipedia is a good starting point to find an initial description of several computer science concepts. For instance, it contains pages that describe, with a good level of detail, all the concepts introduced in this paragraph, i.e. the tree

Silvio Peroni, Francesca Tomasi     Approaching Digital Humanities at university:
A cultural challengeSimilarly, great care is taken about the organisation of the courses from both theoretical and practical points of view. The goal is to put all the theoretical concepts introduced into a pragmatic context by using real-case examples from existing projects or specific practical hands-on sessions studied to work on real scenarios.

Such a practical endeavour is complemented by group projects, representing the means to get the final score of the courses. Indeed, one of the learning outcomes of the entire degree is to enable students to acquire skills for working in groups, mimicking, at a smaller scale, what students will find in the DH market after their graduation. Indeed, a digital humanist is not a person working in an ivory tower (not anymore, at least). Instead, they have to deal with other colleagues – working collaboratively on the same project – and, in particular, with society at large, since more and more researchers are asked to address and identify how the research affects, in a way or another, citizens and communities. For instance, this is particularly true for those working with museums that aim at being 'in the service of society' for increasing the 'participation of communities, offering varied experiences for education, enjoyment, reflection and knowledge sharing',[14] often (today) via the adoption of digital technologies for enabling new narratives and interactions.

Together with this approach to research projects as the final output for demonstrating the acquisition of competencies and skills, and the project-oriented paradigm, based on the collaboration among a group of students, another important point is the consciousness of the whole lifecycle of digital objects in working on data, information and knowledge. Creation, representation, manipulation and dissemination are the activities or actions of a general workflow that is required to keep in mind in each project.

Recently, this master's path has been accompanied by a new PhD programme, CHeDE, starting from the academic year 2022/2023 (cycle XXXVIII). The original idea of this programme was to broaden the coverage of the cultural offer by involving twelve different departments: *Architecture*, *Classical Philology and Italian Studies*, *Computer Science and Engineering*, *Cultural Heritage*, *Economics*, *Education Studies*, *History and Culture*, *Interpreting and Translation*, *Modern Languages, Literatures and Cultures*, *Psychology*, *Philosophy and Communication Studies*, *The Arts*.

---

(https://en.wikipedia.org/wiki/Tree_(data_structure)) and queue (https://en.wikipedia.org/wiki/Queue_(abstract_data_type)) data structures, the FIFO access strategy (https://en.wikipedia.org/wiki/FIFO_(computing_and_electronics)), and the rationale behind the recursion (https://en.wikipedia.org/wiki/Recursion_(computer_science)). However, a more in-depth description of these concepts is provided in several introductory book. The material available at https://comp-think.github.io is a good starting point for deepen in the topic.

[14] Committee for the Museum Definition – ICOM Define, *Final Report (2020-2022)*, Paris, France: ICOM, 2022. https://icom.museum/en/resources/standards-guidelines/museum-definition/

The PhD aims to encourage the growth and development of new specialised skills in which digital tools produce renewed interpretative and analytical approaches for a new and broader valorisation, dissemination and public use of the heritage. After three years of work, the PhDs should have an international and interdisciplinary profile for working at superintendencies, museums, libraries, archives, agencies and organisations related to the study, management and conservation of cultural heritage, private companies that work in cultural heritage and, of course, university research. With this PhD, we would like to introduce a new consciousness in dealing with heritage artefacts resulting from a hermeneutical approach to our cultural tradition.[15]

**Research in Digital Humanities**

The teachings introduced in the previous section accompany and complement the research themes addressed by the /DH.arc. Members of the centre work tightly with, and in, the DHDK master's course and the CHeDE PhD programme to offer teaching that is as close as possible to the current research landscape in the area. The main aim of the /DH.arc is to be a hub for researchers and agencies currently working in the field of DH on innovative projects in the Humanities, and to connect students, researchers, IT staff, and professors affiliated with different departments at the University of Bologna to conduct research together on the DH landscape.

   Among all the DH research topics, the /DH.arc has run several projects in the past years on building specific technical and domain-based expertise in the context of the cultural heritage landscape.[16] A good part of the centre's activity is dedicated to institutional projects aiming at digitising and creating the related metadata for a large set of cultural heritage items to be exhibited via digital libraries. Indeed, the Digital Library of the Department of Classical Philology and Italian Studies (https://dl.ficlit.unibo.it), based on Omeka S (https://omeka.org/s/) and IIIF technologies (https://iiif.io/), is one of the primary efforts published during the past five years as one of the outcomes of the project related to the Departments of Excellence initiative by the Italian Ministry of Education, Universities and Research. The main effort of this activity is to deal with the creation of a new model for a Digital Library, based on Semantic Web technologies, for the production of a knowledge graph of our heritage, represented by data coming from the cultural objects preserved by our Department of Classical Philology and Italian Studies (mostly archival documents and ancient books).

---

[15] For a broader introduction on studies bringing Digital Humanities methodologies, techniques and tools in the context of art historical practices see Kathryn Brown, *The Routledge Companion to Digital Humanities and Art History*, London, United Kingdom: Routledge, 2022.

[16] A full list of the /DH.arc projects can be found at https://centri.unibo.it/dharc/en/research/projects-at-dh-arc.

The study of literary texts is probably one of the first seeds of the entire research work we have dealt with within the centre since the beginning.[17] In particular, recently, we have built on a TEI/XML digital edition of Paolo Bufalini's notebook (http://projects.dharc.unibo.it/bufalini-notebook/) to extend and enrich its content via RDF technologies,[18] with additional data extracted from external sources.[19] In particular, the work has been focused on three different dimensions, i.e. data modelling (the identification of existing ontologies for creating the new data), RDF publication of the new data (via named graphs,[20] and nanopublications),[21] and the development of web applications for serving and exploiting RDF data to build data-driven visualisations. This experience has led to another project, called Linked Data for TEI (LIFT, https://projects.dharc.unibo.it/lift/),[22] strictly connected with the need to move from traditional XML/TEI files to RDF datasets but also to reply to an issue that the community often declare as a crucial point in adopting semantic technologies for managing data.

Another critical aspect of the centre's expertise concerns creating and managing knowledge graphs describing cultural heritage items and their contextual information. An example of this kind of work is the project Zeri & LODE (http://data.fondazionezeri.unibo.it/), which aimed at exposing the vastness of one of the most important collections of European cultural heritage, the Zeri Photo Archive, as Linked Open Data (LOD).[23] The work of this project concerned the development of two new ontologies for describing photographs and the related works of art depicted in such photos to map the descriptive elements used in the current Zeri Photo Archive catalogue into RDF. All these elements have been made available via an RDF triplestore that can be queried through SPARQL (i.e. a query language for RDF data).[24]

---

[17] Related to this field, on the /DH.arc website see also: Edizione Nazionale delle opere di Aldo Moro, Philoeditor, Semantic Digital Edition of Vespasiano da Bisticci's Letters, Vasto.

[18] Richard Cyganiak, David Wood, & Markus Krötzsch, *RDF 1.1 Concepts and Abstract Syntax*, Cambridge, MA, United States of America: World Wide Web Consortium, 2014. https://www.w3.org/TR/rdf11-concepts/

[19] Marilena Daquino, Francesca Giovannetti, & Francesca Tomasi. 'Linked Data per le edizioni scientifiche digitali. Il workflow di pubblicazione dell'edizione semantica del quaderno di appunti di Paolo Bufalini', *Umanistica Digitale*, 7, December 2019. https://doi.org/10.6092/ISSN.2532-8816/9091

[20] Jeremy J. Carroll, Christian Bizer, Pat Hayes, & Patrick Stickler, 'Named graphs', *Journal of Web Semantics*, 3: 4, 247–267, December 2005. https://doi.org/10.1016/j.websem.2005.09.001

[21] Paul Groth, Andrew Gibson, & Jan Velterop, 'The anatomy of a nanopublication', *Information Services & Use*, 30: 1–2, 51–56, September 2010. https://doi.org/10.3233/ISU-2010-0613

[22] Francesca Giovannetti, & Francesca Tomasi, 'Linked data from TEI (LIFT): A Teaching Tool for TEI to Linked Data Transformation', *Digital Humanities Quarterly*, 16: 2, 2022. http://www.digitalhumanities.org/dhq/vol/16/2/000605/000605.html

[23] Marilena Daquino, Francesca Mambelli, Silvio Peroni, Francesca Tomasi, & Fabio Vitali, 'Enhancing Semantic Expressivity in the Cultural Heritage Domain: Exposing the Zeri Photo Archive as Linked Open Data', *Journal on Computing and Cultural Heritage*, 10: 4, 1–21, July 2017. https://doi.org/10.1145/3051487

Strictly related to this project is the whole of the activities connected with the knowledge organisation field, applied to historical documents and works of art. Starting from the need to represent the interpretation act as the result of the activity of the scholar, we developed HiCO (http://purl.org/emmedi/hico) as the ontology for the description of the hermeneutical approach of scholars in reading and understanding data.[25] From HiCO, we have proposed a Digital Hermeneutics theory,[26] defining a model to identify the different levels of data and documents description. An example of this application is MythLOD (https://dharc-org.github.io/mythlod/).[27] The project is an experiment or revalorisation of a digital collection through Linked Open Data format, fostering Semantic Web technologies. It focuses on the formal representation of experts' analysis when associating artworks (and their interpretation) with literary sources. Additionally, the layered approach to knowledge organisation allows one to represent artworks and their descriptive metadata along with interpretations and contextual information.

Since exposing data within a knowledge graph is one of the activities performed in several projects involving members of the /DH.arc, a particular focus is always given to provenance information: records that describe 'the people, institutions, entities, and activities involved in producing, influencing, or delivering a piece of data or a thing' that 'can be used to form assessments about its quality, reliability or trustworthiness'.[28] One of the efforts currently being pursued is to work towards increasing the perceived trustworthiness of data is accompanying such provenance information with *data* change tracking. Some initial ideas have been tested in the context of OpenCitations (https://opencitations.net), an independent not-for-profit infrastructure organisation dedicated to publishing open bibliographic and citation data.[29] The work done in this context, finalised within the frame of a Horizon 2020 project (OpenAIRE-Nexus, https://www.openaire.eu/openaire-nexus-project), resulted in implementing a Python library for enabling the creation of provenance-aware data with change tracking.[30]

---

[24] Steve Harris, & Andy Seaborne, *SPARQL 1.1 Query Language*, Cambridge, MA, United States of America: World Wide Web Consortium, 2013. https://www.w3.org/TR/sparql11-query/

[25] Marilena Daquino, & Francesca Tomasi, 'Historical Context (HiCo): a conceptual model for describing context information of cultural heritage objects' in Emmanouel Garoufallou, Richard J. Hartley, Panorea Gaitanou, eds, *Metadata and Semantic Research*, Cham, Switzerland: Springer 2015 (424–436). https://doi.org/10.1007/978-3-319-24129-6_37

[26] Marilena Daquino, Valentina Pasqual, & Francesca Tomasi, 'Knowledge Representation of digital Hermeneutics of archival and literary Sources', *JLIS.it*, 11: 3, September 2020. https://doi.org/10.4403/jlis.it-12642

[27] Valentina Pasqual, & Francesca Tomasi, 'Linked open data per la valorizzazione di collezioni culturali: il dataset mythLOD', *AIB studi*, 62: 1, 149–168, May 2022. https://doi.org/10.2426/aibstudi-13301

[28] Luc Moreau, & Paolo Missier, *PROV-DM: The PROV Data Model*, Cambridge, MA, United States of America: World Wide Web Consortium, 2013. https://www.w3.org/TR/prov-dm/

Even if all these competencies have been built primarily by working on the textual and bibliographic representation of information, we have had the chance to apply them in other cultural heritage contexts, such as museums. Indeed, in the past three years, an ongoing Horizon 2020 project – SPICE, Social Cohesion, Participation and Inclusion through Cultural Engagement (https://spice-h2020.eu/) – has taken up a great deal of research time. The project includes five case studies involving museums from different countries in devising approaches for *citizen curation*, i.e. using state-of-the-art technologies to develop tools 'for producing, collecting, interpreting, and archiving people's responses to cultural objects, with the aim of favouring the emergence of multiple, sometimes conflicting, viewpoints and motivating the users and memory institutions to reflect upon them'.[31] For instance, one of the results of this project supporting the implementation of a citizen curation methodology has been the implementation of a mechanism to enable a programmer to query and interchange, using SPARQL as a unique interface, museums metadata that can be exposed in various formats, such as CSV, JSON, XML, etc.[32]

Finally, as researchers in the field, the very nature of DH itself is often investigate, trying to understand characterising aspects, borders, and relations with other disciplines. In particular, recently, analysis of DH using a quantitative approach based on DH publications and citations to find which disciplines are closer to DH research and to what extent was undertaken.[33] This work also resulted in the production of a secondary resource with an undebatable utility for the DH community, i.e. an updated list of DH journals (https://dhjournals.github.io/list/).

---

[29] Silvio Peroni, & David Shotton, 'OpenCitations, an infrastructure organization for open scholarship', *Quantitative Science Studies*, 1: 1, 428–444, February 2020. https://doi.org/10.1162/qss_a_00023

[30] Simone Persiani, Marilena Daquino, & Silvio Peroni, 'A Programming Interface for Creating Data According to the SPAR Ontologies and the OpenCitations Data Model' in Paul Groth, Maria-Esther Vidal, Fabian Suchanek, Pedro Szekley, Pavan Kapanipathi, Catia Pesquita, Hala Skaf-Molli, Minna Tamper, eds, *The Semantic Web*, Cham, Switzerland: Springer 2022 (305–322). https://doi.org/10.1007/978-3-031-06981-9_18

[31] Enrico Daga, Luigi Asprino, Rossana Damiano, Marilena Daquino, Belen Diaz Agudo, Aldo Gangemi, Tsvi Kuflik, Antonio Lieto, Mark Maguire, Anna Maria Marras, Delfina Pandiani, Paul Mulholland, Silvio Peroni, Sofia Pescarin, & Alan Wecker, 'Integrating Citizen Experiences in Cultural Heritage Archives: Requirements, State of the Art, and Challenges', *Journal on Computing and Cultural Heritage*, 15: 1, 1–35, January 2022. https://doi.org/10.1145/3477599

[32] Enrico Daga, Luigi Asprino, Paul Mulholland, & Aldo Gangemi, 'Facade-X: An Opinionated Approach to SPARQL Anything' in Mehwish Alam, Paul Groth, Victor de Boer, Tassilo Pellegrini, Harshvardhan J. Pandit, Elena Montiel, Victor Rodríguez Doncel, Barbara McGillivray, Albert Meroño-Peñuela, eds, *Further with Knowledge Graphs*, Amsterdam, The Netherlands: IOS Press 2021 (58–73). https://doi.org/10.3233/SSW210035

**Conclusions**

The experience of the DHDK master's degree allowed us to reflect on a highly cross-disciplinary approach, with mutual interaction between Humanities and Informatics, as well as conceptual contributions from the law for open data and copyright, business strategy and entrepreneurship, communication and social strategy. The course aims at starting from the notion of working inside and for a project: the main point is to focus on the principle of doing research applying DH methods and models to realise scholarly-based digital objects, environments and tools. Blending knowledge to create a new consciousness is the goal of the training provided during the study. Learning is intended as the ability to reason critically about such knowledge to yield innovative and original research results.

In this vision, the cultural shift to Informatics is crucial. Introducing useful digital and computational tools to students in DH without presenting, in advance, the theoretical backgrounds behind them means obliging such students to work without having complete control of what they can do and how they can act. Metaphorically, it is like driving a car without knowing traffic laws and signs.

In the PhD, ideally, this experience is enhanced by opening the view to other domains in the humanities (i.e. archaeology, anthropology, architecture, psychology), but again trying to apply critical/analytical models, methods and approaches to reason on problems to be represented in a formal dimension. Seen as a way to transform reading of the observed reality into something accessible, manageable, and understandable, the digital can function only when the theories surrounding techniques are comprehended entirely.

This approach in education is the result of the research activities undertaken in particular fields that is able to support the overarching research vision: knowledge organisation, digital philology, computational linguistics, knowledge representation, semantic web publishing, data science, and open science. These are the building blocks for a shared national model for the DH.

As a conclusion: research feeds teaching, but teaching is the landscape for the education of the future generations of digital humanists.

---

[33] Gianmarco Spinaci, Giovanni Colavizza, & Silvio Peroni, 'A map of Digital Humanities research across bibliographic data sources', *Digital Scholarship in the Humanities*, 37: 4, 1254–1268, April 2022. https://doi.org/10.1093/llc/fqac016

Silvio Peroni – silvio.peroni@unibo.it

Francesca Tomasi – francesca.tomasi@unibo.it